%% file: QSO.tex
\documentclass[prl,superscriptaddress,showpacs,showkeys,twocolumn,floatfix,nofootinbib]{revtex4}
\usepackage{amsmath,amsfonts,graphics} \input{abbreviations}

\begin{document}
\title{Quantifying Self-Organization with Optimal Predictors}
\author{Cosma Rohilla Shalizi}
\email{cshalizi@umich.edu}
\affiliation{Center for the Study of Complex Systems, University of Michigan,
  Ann Arbor, MI 48109}
\thanks{Supported by a grant from the James S. McDonnell Foundation.}
\author{Kristina Lisa Shalizi}
\email{kshalizi@umich.edu}
\affiliation{Statistics Department, University of Michigan, Ann Arbor, MI
  48109}
\author{Robert Haslinger}
\email{robhh@nmr.mgh.harvard.edu}
\affiliation{MGH-NMR Center, Department of Radiology, Massachusetts General
  Hospital, Charlestown, MA 02129}
\affiliation{Center for Nonlinear Studies, Los Alamos National Laboratory,Los
  Alamos, NM 87545}
\thanks{Supported by DR Project 200153, and the Department of Energy, under
  contract W-7405-ENG-36.}

\begin{abstract}
Despite broad interest in self-organizing systems, there are few quantitative,
experimentally-applicable criteria for self-organization.  The existing
criteria all give counter-intuitive results for important cases.  In this
Letter, we propose a new criterion, namely an internally-generated increase in
the statistical complexity, the amount of information required for optimal
prediction of the system's dynamics.  We precisely define this complexity for
spatially-extended dynamical systems, using the probabilistic ideas of mutual
information and minimal sufficient statistics.  This leads to a general method
for predicting such systems, and a simple algorithm for estimating statistical
complexity.  The results of applying this algorithm to a class of models of
excitable media (cyclic cellular automata) strongly support our proposal.
\end{abstract}


\maketitle

The term ``self-organization'' was coined in the 1940s \cite{Ashby-1947} to
label processes in which systems become more highly organized over time,
without being ordered by outside agents or by external programs.  It has become
one of the leading concepts of nonlinear science, without ever having been
properly defined.  The prevailing ``I know it when I see it'' standard prevents
the development of a {\em theory} of self-organization.  Thus some say that
``self-organizing'' implies ``dissipative''
\cite{Nicolis-Prigogine-self-organization}, and others that they can exhibit
reversible self-organization
\cite{Raissa-DLA,DES-on-self-org-as-refrigeration}, and no one knows if both
groups are talking about the same idea.

A definition of self-organization should be mathematically precise, so we can
build theories around it, and experimentally applicable, so we can use
empirical data to say whether something self-organizes.  The goal of such a
definition should be both to match our informal notions in easy cases, where
intuition is clear and consensual, and to extend unambiguously to intuitively
hard or disputed cases.  If our informal notions allow for comparative, ``more
than'' judgments, a formalization should match those, too.  Generally there are
many ways to formalize a single concept, and competing formalizations must be
judged by their scientific fruitfulness; differing formalizations may be
appropriate in different contexts.  (For more on such methodological issues,
see \cite{Quine-logical}.)

We believe we have a formal criterion for self-organization that meets the key
requirements.  It is precise, unambiguous, and operational.  We check its
conformity with intuition against cellular automata, specifically cyclic
cellular automata (CA).  They are ideal test cases: their dynamics are
completely known (because we specify them) and can easily be simulated exactly.
They are reasonable qualitative models of excitable media, and there is an
analytical theory \cite{Fisch-Gravner-Griffeath-threshold-range} of the
patterns they form.  We show that our definition works, at least in this case.
Two of us discussed preliminary work in \cite{QSO-in-FN03}; here we present the
(concurring) results of larger, more extensive simulations.\footnote{Strictly
  speaking, we quantify system {\em organization}.  In isolated systems, as in
  our simulations, this is necessarily {\em self}-organization.  Distinguishing
  self- from external organization in systems receiving structured input is
  tricky; we discuss some possible approaches below.\\ In any case, our subject
  is distinct from ``self-organized criticality''
  \cite{Bak-Tang-and-Weisenfield}, a term labeling non-equilibrium systems
  whose attractors show power-law fluctuations and long-range correlations.  We
  plan to address whether such systems are self-organizing in our sense in
  future work.}

\vspace{-7mm}
\textsc{Measuring Organization} Few attempts have been made to measure
self-organization quantitatively.  Thermodynamic entropy is an obvious measure
of organization for physicists, and several works claim to measure
self-organization by finding spontaneous declines in entropy
\cite{Wolfram-stat-mech-CA,Krepeau-Isaacson-spectral-entropy-of-self-org,%
  Klimontovich}.  But thermodynamic entropy is a bad measure of organization in
complex systems
\cite{Fox-energy-evolution,Sewell-emergent-macrophysics,Badii-Politi}.  Entropy
is proportional to the logarithm of the accessible volume in phase space, which
has no necessary connection to any kind of organization.  Thus low-temperature
states of Ising systems or Fermi fluids have very low entropy, but no
discernible organization \cite{Sewell-emergent-macrophysics}.  Biological
organisms are {\em never} in pure, low-entropy states, but {\em are} organized,
if anything is.  Some kinds of biological self-organization are, in fact,
thermodynamically driven by {\em increasing} entropy
\cite{Fox-energy-evolution,Privalov-Gill}.

\vspace{-6mm}
After ``fall in entropy'', the leading idea on how to measure
self-organization, advanced in \cite{Bennett-dissipation}, is a rise in
complexity.  While there are many proposed measures of physical complexity, the
general view is that complex phenomena are ones which cannot be described
concisely {\em and} accurately (see \cite{Badii-Politi} for a general survey).
Most proposals use algorithmic descriptions, and are limited by inherent
uncomputability.  Here we take a stochastic point of view, aiming to {\em
  statistically} describe {\em ensembles} of configurations.  We follow
Grassberger \cite{Grassberger-1986} in defining the complexity of a process as
the least amount of information about its state needed for maximally accurate
prediction.  Crutchfield and Young \cite{Inferring-stat-compl} extended this
concept, by giving operational definitions of ``maximally accurate prediction''
and ``state''.

The Grassberger-Crutchfield-Young ``statistical complexity'', $C$, is the
information content of the minimal sufficient statistic for predicting the
process's future \cite{CMPPSS}.  In thermodynamic settings, this is the amount
of information a full set of macrovariables contains about the system's
microscopic state \cite{What-is-a-macrostate}.  We now sketch the formalism
allowing us to use statistical complexity to characterize spatially-extended
dynamical systems of arbitrary dimension, after
\cite{CRS-prediction-on-networks}.

Let $x(\vec{r},t)$ be an $n+1$D field, possibly stochastic, in which
interactions between different space-time points propagate at speed $c$.  As in
\cite{Parlitz-Merkwith-local-states}, define the {\em past light cone} of the
space-time point $(\vec{r}, t)$ as all points which could influence
$x(\vec{r},t)$, i.e., all points $(\vec{q}, u)$ where $u < t$ and $||\vec{q} -
\vec{r}|| \leq c(t-u)$.  The {\em future light cone} of $(\vec{r},t)$ is the
set of all points which could be influenced by what happens at $(\vec{r},t)$.
$\localpast(\vec{r},t)$ is the configuration of the field in the past light
cone, and $\localfuture(\vec{r},t)$ the same for the future light cone.  The
distribution of future light cone configurations, given the configuration in
the past, is $\Prob(\localfuture|\localpast)$.

Any function $\eta$ of $\localpast$ defines a {\em local statistic}.  It
summarizes the influence of all the space-time points which could affect what
happens at $(\vec{r},t)$.  Such local statistics should tell us something about
``what comes next,'' which is $\localfuture$.
(\cite{CRS-prediction-on-networks} explains why we must use {\em local}
predictors, and the advantages of basing them on light cones, as first
suggested by \cite{Parlitz-Merkwith-local-states}.)  Information theory lets us
quantify how informative different statistics are.

The information about variable $x$ in variable $y$ is $I[x;y]$,
\begin{eqnarray}
I[x;y] & \equiv & \left\langle \log_2{\frac{\Prob(x,y)}{\Prob(x)\Prob(y)}}
\right\rangle
\end{eqnarray}
where $\Prob(x,y)$ is joint probability, $\Prob(x)$ is marginal probability,
and $\langle \cdot \rangle$ is expectation
\cite{Kullback-info-theory-and-stats}.  The information a statistic $\eta$
conveys about the future is $I[\localfuture;\eta(\localpast)]$.  A statistic is
{\em sufficient} if it is as informative as possible
\cite{Kullback-info-theory-and-stats}, here if and only if
$I[\localfuture;\eta(\localpast)] = I[\localfuture;\localpast]$.  This is the
same \cite{Kullback-info-theory-and-stats} as requiring that
$\Prob(\localfuture|\eta(\localpast)) = \Prob(\localfuture|\localpast)$.  A
sufficient statistic retains all the predictive information in the data.
Decision theory \cite{Blackwell-Girshick} tells us that maximally accurate and
precise prediction needs {\em only} a sufficient statistic, not the original
data; in fact, any predictor which does not use a sufficient statistic can be
replaced by a superior one which does.  Since we want {\em optimal} prediction,
we confine ourselves to sufficient statistics.

If we use a sufficient statistic $\eta$ for prediction, we must describe or
encode it.  Since $\eta(\localpast)$ is a function of $\localpast$, this
encoding takes $I[\eta(\localpast);\localpast]$ bits.  If knowing $\eta_1$ lets
us compute $\eta_2$, which is also sufficient, then $\eta_2$ is a more concise
summary, and $I[\eta_1(\localpast);\localpast] \geq
I[\eta_2(\localpast);\localpast]$.  A {\em minimal sufficient statistic}
\cite{Kullback-info-theory-and-stats} can be computed from any other sufficient
statistic.  We now construct one.

Take two past light cone configurations, $\localpast_1$ and $\localpast_2$.
Each has some conditional distribution over future light cone configurations,
$\Prob(\localfuture|\localpast_1)$ and $\Prob(\localfuture|\localpast_2)$
respectively.  The two past configurations are equivalent, $\localpast_1 \sim
\localpast_2$, if those conditional distributions are equal.  The set of
configurations equivalent to $\localpast$ is $[\localpast]$. Our statistic is
the function which maps past configurations to their equivalence classes:
\begin{equation}
\epsilon(\localpast)  \equiv [\localpast]
 =  \left\{\localpastprime ~: ~\Prob(\localfuture|\localpastprime) = \Prob(\localfuture|\localpast)\right\}
\end{equation}
Clearly, $\Prob(\localfuture|\epsilon(\localpast)) =
\Prob(\localfuture|\localpast)$, and so $I[\localfuture;\epsilon(\localpast)] =
I[\localfuture;\localpast]$, making $\epsilon$ a sufficient statistic.  The
equivalence classes, the values $\epsilon$ can take, are the {\em causal
  states} \cite{CRS-prediction-on-networks,Inferring-stat-compl,CMPPSS,%
  What-is-a-macrostate}.  Each causal state is a set of specific past
light-cones, and all the cones it contains are equivalent, predicting the same
possible futures with the same probabilities.  Thus there is no advantage to
subdividing the causal states, which are the coarsest set of predictively
sufficient states.

For any sufficient statistic $\eta$, $\Prob(\localfuture|\localpast) =
\Prob(\localfuture|\eta(\localpast))$.  So if $\eta(\localpast_1) =
\eta(\localpast_2)$, then $\Prob(\localfuture|\localpast_1) =
\Prob(\localfuture|\localpast_2)$, and the two pasts belong to the same causal
state.  Since we can get the causal state from $\eta(\localpast)$, we can use
the latter to compute $\epsilon(\localpast)$.  Thus, $\epsilon$ is minimal.
Moreover, $\epsilon$ is the {\em unique} minimal sufficient statistic
\cite{CRS-prediction-on-networks}: any other just relabels the same states.

Because $\epsilon$ is minimal, $I[\epsilon(\localpast);\localpast] \leq
I[\eta(\localpast);\localpast]$, for any other sufficient statistic $\eta$.
Thus we can speak objectively about the minimal amount of information needed to
predict the system, which is how much information about the past of the system
is relevant to predicting its own dynamics.  This quantity,
$I[\epsilon(\localpast);\localpast]$, is a characteristic of the system, and
not of any particular model.  We define the {\em statistical complexity} as
\begin{eqnarray}
C & \equiv & I[\epsilon(\localpast);\localpast]
\end{eqnarray}
$C$ is the amount of information required to describe the behavior at that
point, and equals the log of the effective number of causal states, i.e., of
different distributions for the future.  Complexity lies between disorder and
order \cite{Badii-Politi,Grassberger-1986,Inferring-stat-compl}, and $C = 0$
both when the field is completely disordered (all values of $x$ are
independent) and completely ordered ($x$ is constant). $C$ grows when the
field's dynamics become more flexible and intricate, and more information is
needed to describe the behavior.

We now sketch an algorithm to recover the causal states from data, and so
estimate $C$.  (\cite{CRS-prediction-on-networks} provides details, including
pseudocode; cf.\ \cite{Parlitz-Merkwith-local-states}.)  At each time $t$, list
the observed past and future light-cone configurations, and put the observed
past configurations in some arbitrary order, $\left\{\localpast_i\right\}$.
(In practice, we must limit how far light-cones extend into the past or
future.)  For each past configuration $\localpast_i$, estimate
$\Prob_t(\localfuture|\localpast_i)$.  We want to estimate the states, which
ideally are groups of past cones with the same conditional distribution over
future cone configurations.  Not knowing the conditional distributions a
priori, we must estimate them from data, and with finitely many samples, such
estimates always have some error.  Thus, we approximate the true causal states
by clusters of past light-cones with similar distributions over future
light-cones; the conditional distribution for a cluster is the weighted mean of
those of its constituent past cones.  Start by assigning the first past,
$\localpast_1$ to the first cluster.  Thereafter, for each $\localpast_i$, go
down the list of existing clusters and check whether
$\Prob_t(\localfuture|\localpast_i)$ differs significantly from each cluster's
distribution, as determined by a fixed-size $\chi^2$ test.  (We used $\alpha =
0.05$ in our simulations below.)  If the discrepancy is insignificant, add
$\localpast_i$ to the first matching cluster, updating the latter's
distribution.  Make a new cluster if $\localpast_i$ does not match any existing
cluster.  Continue until every $\localpast_i$ is assigned to some cluster.  The
clusters are then the estimated causal states at time $t$.  Finally, obtain the
probabilities of the different causal states from the empirical probabilities
of their constituent past configurations, and calculate $C(t)$.  This procedure
converges on the correct causal states as it gets more data, independent of the
order of presentation of the past light-cones, the ordering of the clusters, or
the size $\alpha$ of the significance test \cite{CRS-prediction-on-networks}.
For finite data, the order of presentation matters, but we finesse this by
randomizing the order.

We say a system has {\em organized} between times $t_1$ and $t_2$ if (I)
$C(t_2) - C(t_1) \equiv \Delta C > 0$.  It has {\em self-organized} if (II)
some of the rise in complexity is not due to external agents.  We can check
condition (I) by estimating $\Delta C$.  We know condition (II) holds for many
systems, because they either have no external inputs (e.g., deterministic CA),
or only unstructured inputs (e.g., chemical pattern-formers exposed to thermal
noise).  For systems with structured input, we need, but lack, a way to say how
much of $\Delta C$ is due to that input.  We could, perhaps, treat this as a
causal inference problem \cite{Pearl-causality}, with $\Delta C$ as the
response variable, and the input as the treatment.  Alternately, we could see
how much $\Delta C$ changes if we replace the input with statistically-similar
noise \cite{Delgado-Sole-collective-induced}.

\textsc{Numerical Experiments and Results} Having developed a quantitative
criterion for self-organization, we now check it experimentally.  Our test
systems are cyclic cellular automata
\cite{Fisch-Gravner-Griffeath-threshold-range} (CCA), which are models of
pattern formation in excitable media \cite{Tyson-Keener-excitable-review}.
Each site in a square lattice has one of $\kappa$ colors.  A cell of color $k$
will change its color to $k+1 \bmod \kappa$ if there are already at least $T$
(``threshold'') cells of that color in its neighborhood, i.e., within a
distance $r$ (``range'') of that cell.  Otherwise, the cell keeps its current
color.  (In normal excitable media, which have a unique quiescent state, the
role of the threshold is slightly different
\cite{Tyson-Keener-excitable-review}.)  All cells update their colors in
parallel.

\begin{figure}
\begin{center}
(a) \resizebox{0.40\columnwidth}{!}{\includegraphics*[4.45in,3.35in]{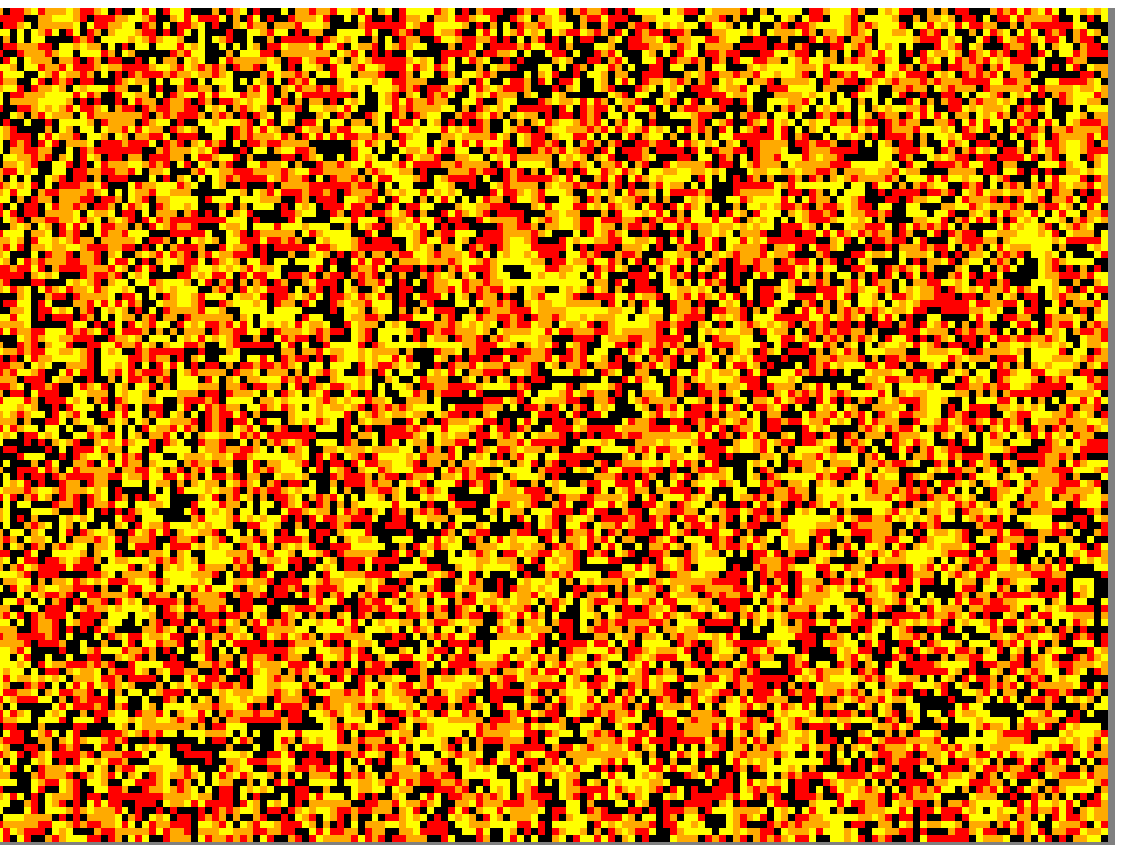}}
(b) \resizebox{0.40\columnwidth}{!}{\includegraphics*[4.45in,3.35in]{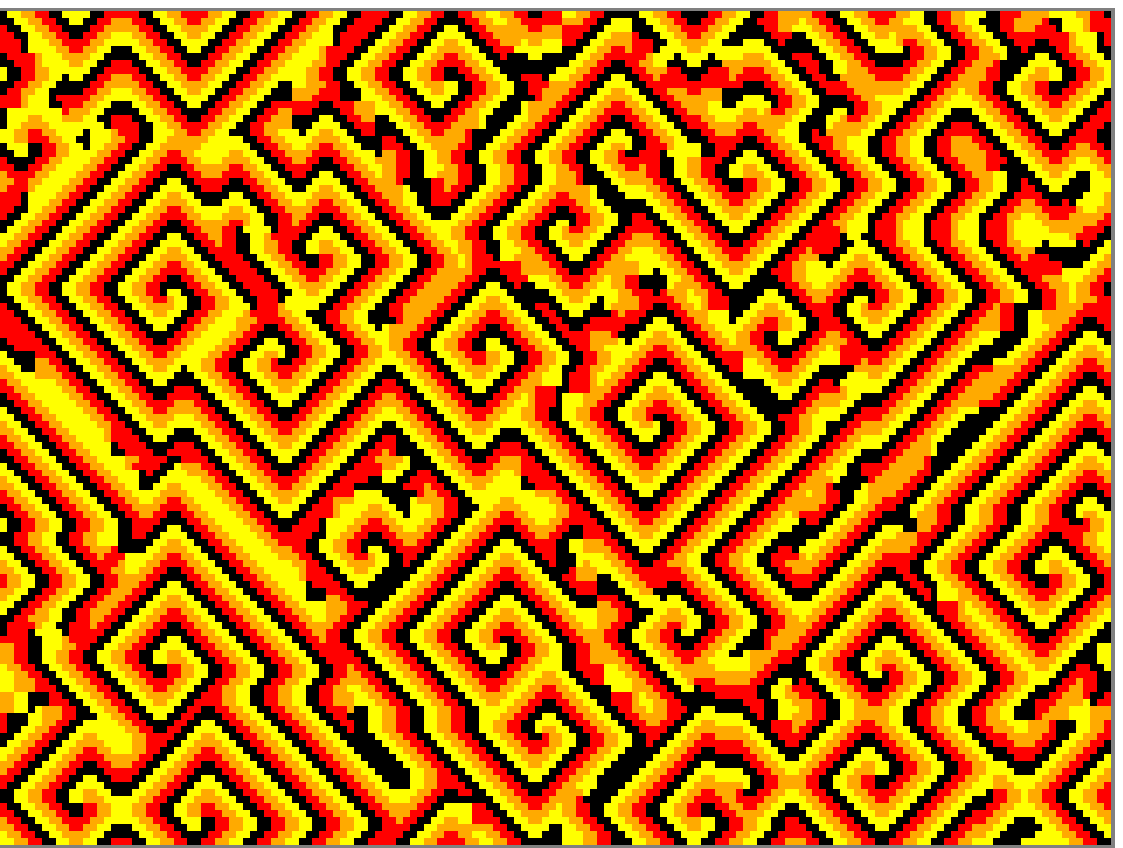}}\\
(c) \resizebox{0.40\columnwidth}{!}{\includegraphics*[4.45in,3.35in]{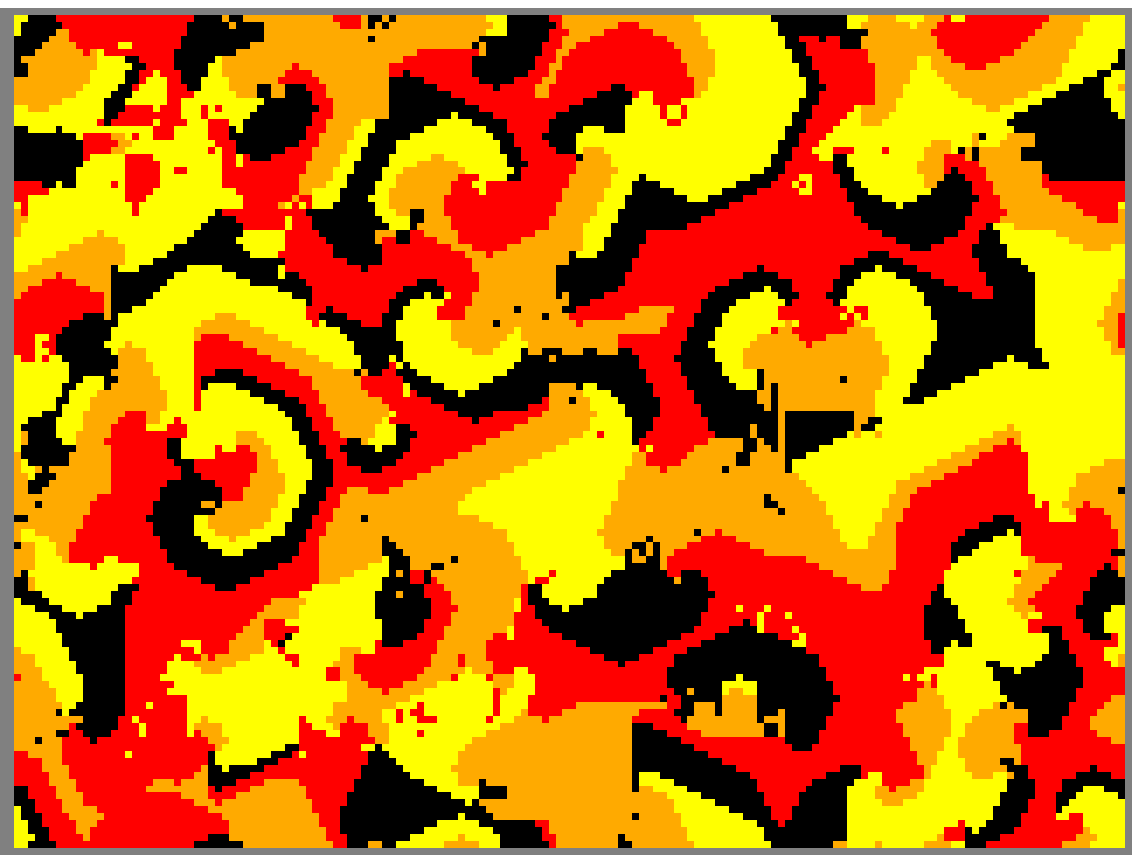}}
(d) \resizebox{0.40\columnwidth}{!}{\includegraphics*[4.45in,3.35in]{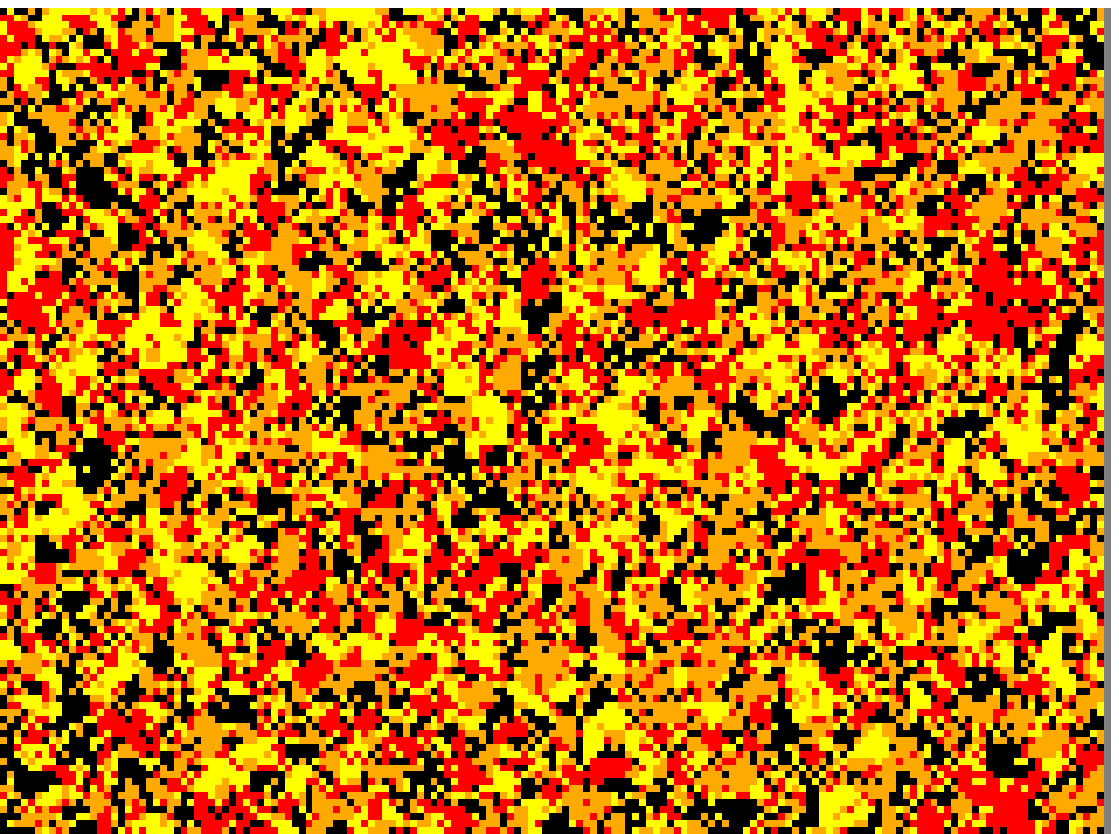}}
\end{center}
\caption{\label{fig:shots} (Color online.)  Phases of the cyclic CA.
  Parameters are as described in the text, started from uniform random initial
  conditions.  Color figures were prepared with \cite{MCell}.  From the top
  left: (a) Local oscillations $(T=1)$, in which the CA oscillates with period
  4, each cell cycling through all colors; (b) Spiral waves $(T=2)$; (c) The
  ``turbulent'' phase $(T=3)$; (d) Fixation with solid color blocks $(T=4)$.}
\vspace{-6mm}
\end{figure}

CCA have three generic long-run behaviors, depending on the ratio of the
threshold to the range.  At high thresholds, CCA form homogeneous blocks of
solid colors, which are completely static (``fixation'').  At very low
thresholds, the entire lattice eventually oscillates periodically; sometimes
rotating spiral waves grow to engulf the entire lattice.  With intermediate
thresholds, incoherent traveling waves form, propagate, collide and disperse;
this, metaphorically, is ``turbulence''. With a range one Moore (box)
neighborhood and $\kappa=4$, the phenomenology is as follows
\cite{Fisch-Gravner-Griffeath-threshold-range} (see Fig.\ \ref{fig:shots}).
$T=1$ and $T=2$ are both locally periodic, but $T=2$ produces spiral waves,
while $T=1$ quenches incoherent local oscillations.  $T=3$ leads to meta-stable
turbulence --- spiral waves can form and entrain the entire CA, but turbulence
can persist indefinitely on finite lattices.  Fixation occurs with $T \geq 4$.
All CCA phases self-organize when started from uniform noise.  (This is best
appreciated by viewing simulations \cite{MCell}.)  By the same intuitive
standard, the fixation phase is less organized than turbulence (which has
dynamic, large-scale spatial structures), which in turn is less organized than
spiral waves (which has more intricate structures).  It is hard to say, by eye,
whether incoherent local oscillations are more or less organized than simple
fixation.  All four regimes lead to stable stationary distributions.  Thus, $C$
should start at zero (reflecting the totally random initial conditions), rise
to a steady value, and stay there.  $T=2$ should have the highest long-run
complexity, followed by $T=3$.

\begin{figure}
\begin{center}
\resizebox{\columnwidth}{!}{\includegraphics{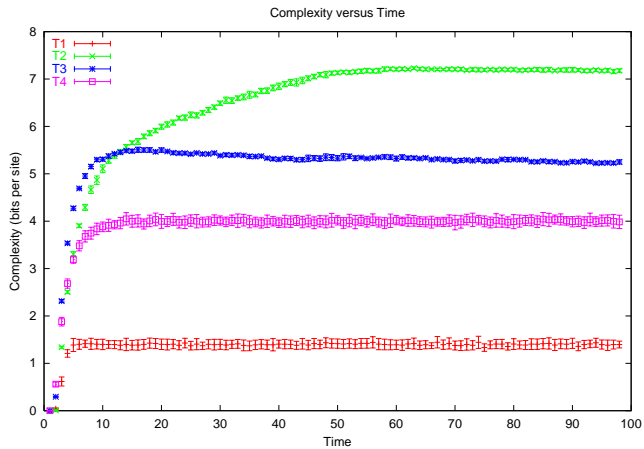}}
\end{center}
\caption{(Color online.)  \label{fig:results} Complexity over time for CCA with
  different thresholds $T$, averaging 30 independent simulations at each value
  of $T$. The $T=2$ curve has the highest asymptote, followed by $T=3$, $T=4$
  and $T=1$.  Error bars: standard error of the complexity.}
\vspace{-5mm}
\end{figure}

We ran $\kappa=4$, $r=1$ CCA on $300 \times 300$ lattices with periodic
boundary conditions, for $T$ from 1 to 4.  Figure \ref{fig:results} shows the
results of applying our proposed measure of self-organization to these
simulations.  We used light-cones extending 1 time-step into both past and
future; longer light-cones did not, here, lead to different states.  The
agreement with expectations is clear.  All four curves climb steadily to
plateaus, leveling off when the distribution of CA configurations become
stationary.  Sampling noise leads to fluctuations around the asymptotic values
\cite{QSO-in-FN03}.  The slight fall in complexity for $T=3$ occurs when
spirals try to form but break up, and their debris limit further spiral
formation.  Additional simulations at different lattice sizes $L$ show the
estimated long-run complexity growing with $L$, approaching a limit as
$O(L^{-1})$.  This rate combines finite-size effects with the negative bias of
our information estimator, which is at least $O(L^{-2})$
\cite{Victor-information-bias}. We hope in the future to precisely determine
both our estimation bias and the finite-size scaling of the complexity.

\textsc{Conclusion} A theory of self-organization should predict when and why
different systems will assume different kinds and degrees of organization.
This will require an adequate characterization of self-organization.  We argue
that ``internally-caused rise in complexity'' works, if we define complexity as
the amount of information needed for optimal statistical prediction.  We can
reliably estimate this statistical complexity from data, and for CCA, the
estimates match intuitive judgments about self-organization.  The methods used
are not limited to CA, but apply to all kinds of discrete random fields,
including ones on complex networks \cite{CRS-prediction-on-networks}.  They
would work equally well on discretized empirical data, e.g., digital movies of
chemical pattern formation experiments.  This is a first step towards a
physical theory of self-organization.

\emph{Acknowledgments} We thank D. Abbott, J. Crutchfield, R. D'Souza,
D. Feldman, D. Griffeath, C. Moore, S. Page, M. Porter, E. Smith,
J. Usinowicz and our referees.

\bibliography{locusts}
\bibliographystyle{apsrev}

\end{document}

%% file: abbreviations.tex
\newcommand{\Prob}	{ {\rm P} }

\newcommand{\localpast}			{ l^{-}}
\newcommand{\localpastprime}		{ \lambda}

\newcommand{\localfuture}		{ {l}^{+}}





%% file: QSO.bbl
\begin{thebibliography}{26}
\expandafter\ifx\csname natexlab\endcsname\relax\def\natexlab#1{#1}\fi
\expandafter\ifx\csname bibnamefont\endcsname\relax
  \def\bibnamefont#1{#1}\fi
\expandafter\ifx\csname bibfnamefont\endcsname\relax
  \def\bibfnamefont#1{#1}\fi
\expandafter\ifx\csname citenamefont\endcsname\relax
  \def\citenamefont#1{#1}\fi
\expandafter\ifx\csname url\endcsname\relax
  \def\url#1{\texttt{#1}}\fi
\expandafter\ifx\csname urlprefix\endcsname\relax\def\urlprefix{URL }\fi
\providecommand{\bibinfo}[2]{#2}
\providecommand{\eprint}[2][]{\url{#2}}

\bibitem[{\citenamefont{Ashby}(1947)}]{Ashby-1947}
  \bibinfo{author}{\bibfnamefont{W.~R.} \bibnamefont{Ashby}},
  \bibinfo{journal}{J. General Psychology} \textbf{\bibinfo{volume}{37}},
  \bibinfo{pages}{125} (\bibinfo{year}{1947}).

\bibitem[{\citenamefont{Nicolis and
  Prigogine}(1977)}]{Nicolis-Prigogine-self-organization}
\bibinfo{author}{\bibfnamefont{G.}~\bibnamefont{Nicolis}} \bibnamefont{and}
  \bibinfo{author}{\bibfnamefont{I.}~\bibnamefont{Prigogine}},
  \emph{\bibinfo{title}{Self-Organization in Nonequilibrium Systems}}
  (\bibinfo{publisher}{Wiley}, \bibinfo{address}{New York},
  \bibinfo{year}{1977}).

\bibitem[{\citenamefont{D'Souza and Margolus}(1999)}]{Raissa-DLA}
\bibinfo{author}{\bibfnamefont{R.~M.} \bibnamefont{D'Souza}} \bibnamefont{and}
  \bibinfo{author}{\bibfnamefont{N.~H.} \bibnamefont{Margolus}},
  \bibinfo{journal}{Phys. Rev. E} \textbf{\bibinfo{volume}{60}},
  \bibinfo{pages}{264} (\bibinfo{year}{1999}),
  \urlprefix\url{cond-mat/9810258}.

\bibitem[{\citenamefont{Smith}(2003)}]{DES-on-self-org-as-refrigeration}
\bibinfo{author}{\bibfnamefont{E.}~\bibnamefont{Smith}},
  \bibinfo{journal}{Phys. Rev. E} \textbf{\bibinfo{volume}{68}},
  \bibinfo{pages}{046114} (\bibinfo{year}{2003}).

\bibitem[{\citenamefont{Quine}(1961)}]{Quine-logical}
  \bibinfo{author}{\bibfnamefont{W.~V.~O.} \bibnamefont{Quine}},
  \emph{\bibinfo{title}{From a Logical Point of View}}
  (\bibinfo{publisher}{Harvard U. P.}, \bibinfo{address}{Cambridge,
    Mass.}, \bibinfo{year}{1953}).

\bibitem[{\citenamefont{Fisch et~al.}(1991)\citenamefont{Fisch, Gravner, and
  Griffeath}}]{Fisch-Gravner-Griffeath-threshold-range}
\bibinfo{author}{\bibfnamefont{R.}~\bibnamefont{Fisch}},
  \bibinfo{author}{\bibfnamefont{J.}~\bibnamefont{Gravner}}, \bibnamefont{and}
  \bibinfo{author}{\bibfnamefont{D.}~\bibnamefont{Griffeath}},
  \bibinfo{journal}{Stat. Comput.} \textbf{\bibinfo{volume}{1}},
  \bibinfo{pages}{23} (\bibinfo{year}{1991}),
  \urlprefix\url{psoup.math.wisc.edu/papers/tr.zip}.

\bibitem[{\citenamefont{Shalizi and Shalizi}(2003)}]{QSO-in-FN03}
  \bibinfo{author}{\bibfnamefont{C.~R.} \bibnamefont{Shalizi}}
  \bibnamefont{and} \bibinfo{author}{\bibfnamefont{K.~L.}
    \bibnamefont{Shalizi}}, in \emph{\bibinfo{booktitle}{Noise in Complex
      Systems and Stochastic Dynamics}}, edited by
  \bibinfo{editor}{\bibfnamefont{L.}~\bibnamefont{Schimansky-Geier} et al.}
  (\bibinfo{publisher}{SPIE}, \bibinfo{address}{Bellingham, Washington},
  \bibinfo{year}{2003}), pp. \bibinfo{pages}{108--117},
    \urlprefix\url{bactra.org/research/FN03.pdf}.

\bibitem[{\citenamefont{Bak et~al.}(1987)\citenamefont{Bak, Tang, and
  Wiesenfeld}}]{Bak-Tang-and-Weisenfield}
\bibinfo{author}{\bibfnamefont{P.}~\bibnamefont{Bak}},
  \bibinfo{author}{\bibfnamefont{C.}~\bibnamefont{Tang}}, \bibnamefont{and}
  \bibinfo{author}{\bibfnamefont{K.}~\bibnamefont{Wiesenfeld}},
  \bibinfo{journal}{Phys. Rev. Lett.} \textbf{\bibinfo{volume}{59}},
  \bibinfo{pages}{381} (\bibinfo{year}{1987}).

\bibitem[{\citenamefont{Wolfram}(1983)}]{Wolfram-stat-mech-CA}
\bibinfo{author}{\bibfnamefont{S.}~\bibnamefont{Wolfram}},
  \bibinfo{journal}{Rev. Mod. Phys.} \textbf{\bibinfo{volume}{55}},
  \bibinfo{pages}{601} (\bibinfo{year}{1983}).

\bibitem[{\citenamefont{Krepeau and
  Isaacson}(1990)}]{Krepeau-Isaacson-spectral-entropy-of-self-org}
\bibinfo{author}{\bibfnamefont{J.~C.} \bibnamefont{Krepeau}} \bibnamefont{and}
  \bibinfo{author}{\bibfnamefont{L.~K.} \bibnamefont{Isaacson}},
  \bibinfo{journal}{J. Noneq. Thermo.}
  \textbf{\bibinfo{volume}{15}}, \bibinfo{pages}{115} (\bibinfo{year}{1990}).

\bibitem[{\citenamefont{Klimontovich}(1991)}]{Klimontovich}
\bibinfo{author}{\bibfnamefont{Y.~L.} \bibnamefont{Klimontovich}},
  \emph{\bibinfo{title}{Turbulent Motion and the Structure of Chaos}}
  (\bibinfo{publisher}{Kluwer}, \bibinfo{address}{Dordrecht},
  \bibinfo{year}{1991}).

\bibitem[{\citenamefont{Fox}(1988)}]{Fox-energy-evolution}
\bibinfo{author}{\bibfnamefont{R.~F.} \bibnamefont{Fox}},
  \emph{\bibinfo{title}{Energy and the Evolution of Life}}
  (\bibinfo{publisher}{Freeman}, \bibinfo{address}{New York},
  \bibinfo{year}{1988}).

\bibitem[{\citenamefont{Sewell}(2002)}]{Sewell-emergent-macrophysics}
\bibinfo{author}{\bibfnamefont{G.~L.} \bibnamefont{Sewell}},
  \emph{\bibinfo{title}{Quantum Mechanics and Its Emergent Macrophysics}}
  (\bibinfo{publisher}{Princeton U. P.},
  \bibinfo{address}{Princeton}, \bibinfo{year}{2002}).

\bibitem[{\citenamefont{Badii and Politi}(1997)}]{Badii-Politi}
\bibinfo{author}{\bibfnamefont{R.}~\bibnamefont{Badii}} \bibnamefont{and}
  \bibinfo{author}{\bibfnamefont{A.}~\bibnamefont{Politi}},
  \emph{\bibinfo{title}{Complexity: Hierarchical Structures and Scaling in
  Physics}} (\bibinfo{publisher}{Cambridge U. P.},
  \bibinfo{address}{Cambridge}, \bibinfo{year}{1997}).

\bibitem[{\citenamefont{Privalov and Gill}(1988)}]{Privalov-Gill}
\bibinfo{author}{\bibfnamefont{P.~L.} \bibnamefont{Privalov}} \bibnamefont{and}
  \bibinfo{author}{\bibfnamefont{S.~J.} \bibnamefont{Gill}},
  \bibinfo{journal}{Adv. Protein Chem.}
  \textbf{\bibinfo{volume}{39}}, \bibinfo{pages}{191} (\bibinfo{year}{1988}).

\bibitem[{\citenamefont{Bennett}(1985)}]{Bennett-dissipation}
\bibinfo{author}{\bibfnamefont{C.~H.} \bibnamefont{Bennett}}, in
  \emph{\bibinfo{booktitle}{Emerging Syntheses in Science}}, edited by
  \bibinfo{editor}{\bibfnamefont{D.}~\bibnamefont{Pines}}
  (\bibinfo{publisher}{Santa Fe Institute}, \bibinfo{address}{Santa Fe, New
  Mexico}, \bibinfo{year}{1985}), pp. \bibinfo{pages}{215--234}.

\bibitem[{\citenamefont{Grassberger}(1986)}]{Grassberger-1986}
\bibinfo{author}{\bibfnamefont{P.}~\bibnamefont{Grassberger}},
  \bibinfo{journal}{Int. J. Theor. Phys.}
  \textbf{\bibinfo{volume}{25}}, \bibinfo{pages}{907} (\bibinfo{year}{1986}).

\bibitem[{\citenamefont{Crutchfield and Young}(1989)}]{Inferring-stat-compl}
\bibinfo{author}{\bibfnamefont{J.~P.} \bibnamefont{Crutchfield}}
  \bibnamefont{and} \bibinfo{author}{\bibfnamefont{K.}~\bibnamefont{Young}},
  \bibinfo{journal}{Phys. Rev. Lett.} \textbf{\bibinfo{volume}{63}},
  \bibinfo{pages}{105} (\bibinfo{year}{1989}).

\bibitem[{\citenamefont{Shalizi and Crutchfield}(2001)}]{CMPPSS}
  \bibinfo{author}{\bibfnamefont{C.~R.} \bibnamefont{Shalizi}}
  \bibnamefont{and} \bibinfo{author}{\bibfnamefont{J.~P.}
    \bibnamefont{Crutchfield}}, \bibinfo{journal}{J. Stat. Phys.}
  \textbf{\bibinfo{volume}{104}}, \bibinfo{pages}{817} (\bibinfo{year}{2001}),
  \urlprefix\url{cond-mat/9907176}.

\bibitem[{\citenamefont{Shalizi and Moore}(2003)}]{What-is-a-macrostate}
  \bibinfo{author}{\bibfnamefont{C.~R.} \bibnamefont{Shalizi}}
  \bibnamefont{and} \bibinfo{author}{\bibfnamefont{C.}~\bibnamefont{Moore}},
  \bibinfo{journal}{Studies Hist. Phil. Mod. Phys.}
  \textbf{\bibinfo{volume}{submitted}} (\bibinfo{year}{2003}),
  \urlprefix\url{cond-mat/0303625}.

\bibitem[{\citenamefont{Shalizi}(2003)}]{CRS-prediction-on-networks}
\bibinfo{author}{\bibfnamefont{C.~R.} \bibnamefont{Shalizi}},
  \bibinfo{journal}{Discrete Math. Theor. Comput. Sci.}
  \textbf{\bibinfo{volume}{AB(DMCS)}}, \bibinfo{pages}{11}
  (\bibinfo{year}{2003}), \urlprefix\url{math.PR/0305160}.

\bibitem[{\citenamefont{Parlitz and
  Merkwirth}(2000)}]{Parlitz-Merkwith-local-states}
\bibinfo{author}{\bibfnamefont{U.}~\bibnamefont{Parlitz}} \bibnamefont{and}
  \bibinfo{author}{\bibfnamefont{C.}~\bibnamefont{Merkwirth}},
  \bibinfo{journal}{Phys. Rev. Lett.} \textbf{\bibinfo{volume}{84}},
  \bibinfo{pages}{1890} (\bibinfo{year}{2000}).

\bibitem[{\citenamefont{Kullback}(1959)}]{Kullback-info-theory-and-stats}
\bibinfo{author}{\bibfnamefont{S.}~\bibnamefont{Kullback}},
  \emph{\bibinfo{title}{Information Theory and Statistics}}
  (\bibinfo{publisher}{Wiley}, \bibinfo{address}{New York},
  \bibinfo{year}{1959}).

\bibitem[{\citenamefont{Blackwell and Girshick}(1954)}]{Blackwell-Girshick}
  \bibinfo{author}{\bibfnamefont{D.}~\bibnamefont{Blackwell}} \bibnamefont{and}
  \bibinfo{author}{\bibfnamefont{M.~A.} \bibnamefont{Girshick}},
  \emph{\bibinfo{title}{Theory of Games and Statistical Decisions}}
  (\bibinfo{publisher}{Wiley}, \bibinfo{address}{New York},
  \bibinfo{year}{1954}).


\bibitem[{\citenamefont{Pearl}(2000)}]{Pearl-causality}
\bibinfo{author}{\bibfnamefont{J.}~\bibnamefont{Pearl}},
  \emph{\bibinfo{title}{Causality: Models, Reasoning, and Inference}}
  (\bibinfo{publisher}{Cambridge U. P.},
  \bibinfo{address}{Cambridge}, \bibinfo{year}{2000}).

\bibitem[{\citenamefont{Delgado and
  Sol{\'e}}(1997)}]{Delgado-Sole-collective-induced}
\bibinfo{author}{\bibfnamefont{J.}~\bibnamefont{Delgado}} \bibnamefont{and}
  \bibinfo{author}{\bibfnamefont{R.~V.} \bibnamefont{Sol{\'e}}},
  \bibinfo{journal}{Physical Review E} \textbf{\bibinfo{volume}{55}},
  \bibinfo{pages}{2338} (\bibinfo{year}{1997}).

\bibitem[{\citenamefont{Tyson and
  Keener}(1988)}]{Tyson-Keener-excitable-review}
\bibinfo{author}{\bibfnamefont{J.~J.} \bibnamefont{Tyson}} \bibnamefont{and}
  \bibinfo{author}{\bibfnamefont{J.~P.} \bibnamefont{Keener}},
  \bibinfo{journal}{Physica D} \textbf{\bibinfo{volume}{32}},
  \bibinfo{pages}{327} (\bibinfo{year}{1988}).

\bibitem[{\citenamefont{W{\'o}jtowicz}(2002)}]{MCell}
\bibinfo{author}{\bibfnamefont{M.}~\bibnamefont{W{\'o}jtowicz}},
  \emph{\bibinfo{title}{Cellebration}}, \bibinfo{howpublished}{Online software}
  (\bibinfo{year}{2002}), \urlprefix\url{psoup.math.wisc.edu/mcell/}.

\bibitem[{\citenamefont{Victor}(2000)}]{Victor-information-bias}
\bibinfo{author}{\bibfnamefont{J.~D.} \bibnamefont{Victor}},
  \bibinfo{journal}{Neural Computation} \textbf{\bibinfo{volume}{12}},
  \bibinfo{pages}{2797} (\bibinfo{year}{2000}).

\end{thebibliography}
